\documentstyle[12pt,world_sci]{article}
\renewcommand{\to}{\rightarrow}

\newcommand{\pv}{{\bf p}}

\begin{document}

\title{{\bf QCD POTENTIAL MODEL FOR LIGHT-HEAVY QUARKONIA
AND THE HEAVY QUARK EFFECTIVE THEORY}}
\author{SURAJ N. GUPTA and JAMES M. JOHNSON\footnote{Presented by James M.
Johnson}\\
{\em Department of Physics, Wayne State University, Detroit, Michigan,
48202}\\}

\maketitle
\setlength{\baselineskip}{2.6ex}

\begin{center}
\parbox{13.0cm}
{\begin{center} ABSTRACT \end{center}
{\small \hspace*{0.3cm}
	We have investigated the spectra of light-heavy quarkonia with the
use of a quantum-chromodynamic potential model
which is similar to that used earlier
for the heavy quarkonia.  An essential feature of our treatment is the
inclusion of the one-loop radiative
corrections to the  quark-antiquark potential,
which contribute significantly to the spin-splittings
among the quarkonium energy levels.
Unlike $c\bar{c}$ and $b\bar{b}$, the potential for a
light-heavy system has a complicated dependence on the light and heavy
quark masses $m$ and $M$, and
it contains a spin-orbit mixing term.  We have
obtained excellent results for the observed energy levels
of $D^0$, $D_s$, $B^0$, and $B_s$, and we are able to
provide predicted results for many unobserved
energy levels.

\hspace*{0.3cm}We have also used our investigation to test the accuracy
of the heavy quark effective theory.  We find that
the heavy quark expansion yields generally good results for the $B^0$
and $B_s$ energy levels provided that $M^{-1}$ and $M^{-1}\ln M$
corrections are taken into account in the quark-antiquark interactions.
It does not, however, provide equally good results for the energy
levels of $D^0$ and $D_s$, which shows that the effective theory can be
applied more accurately to the $b$ quark than the $c$ quark.
}}
\end{center}

\section{Introduction}

	The light-heavy quarkonia $D$, $D_s$, $B$, and $B_s$ are
at present of much experimental and theoretical interest, and their exploration
is necessary for our understanding of the strong as well as the
electroweak interactions.  We shall here investigate
the spectra of light-heavy quarkonia with the use
of a quantum chromodynamic model similar to the highly successful
model used earlier for the heavy
quarkonia $c\bar{c}$ and $b\bar{b}$ \cite{ccbar}.  The complexity
of the model is necessarily enhanced for a light-heavy system because
the potential has a complicated dependence on the light
and heavy quark masses $m$ and $M$, and it contains a spin-orbit
mixing term.

\section{Light-heavy Quarkonium Spectra}

	Our treatment for the light-heavy quarkonia is similar to that
for $c\bar{c}$ and $b\bar{b}$ except for the complications arising from the
difference in the quark and antiquark masses.  Thus, our model
is based on the Hamiltonian
\begin{equation}
H=H_0+V_p+V_c,\quad
H_0=(m^2+\pv^2)^{1/2}+(M^2+\pv^2)^{1/2}
\end{equation}
where $V_p$ and $V_c$ are nonsingular
quasistatic perturbative and confining potentials.  Since our
potentials are nonsingular, we are able to avoid the use of an illegitimate
perturbative treatment.

	The experimental and theoretical results for the energy levels
of the light-heavy quarkonia $D^0$, $B^0$, $D_s$, and $B_s$, together
with the ${}^3P^\prime_1$-${}^1P^\prime_1$ mixing angles
arising from the spin-orbit mixing terms, are given in
Table~\ref{d0}.
For experimental data we have relied on the Particle Data Group\cite{pdg}
except that we have used
the more recent results from the CLEO collaboration\cite{cleo1,cleo10}
for $D_1^0$, $D_2^{\star0}$, and $D_{s2}$ and from the
CDF collaboration\cite{cdf} for $B_s$.
In this table, one set of theoretical results corresponds to the
direct use of our model, while the other two sets are obtained
by means of heavy quark expansions of our potentials
to test the accuracy of the heavy quark effective theory
with the inclusion of the $M^{-1}$ and $M^{-1}\ln M$ corrections as
well as without these corrections.

	We expect the dynamics of a light-heavy system
to be primarily dependent on the light quark.  Therefore, our potential
paramenters for $D^0$ and $B^0$ are the same except for the
difference in the $c$ and $b$ quark masses.
We have also ensured that the parameters for $D_s$ and $B_s$ are related to
those for $D^0$ and $B^0$ through quantum
chromodynamic transformation relations.

\section{Conclusion}

	We have obtained excellent results for the observed energy levels
of $D^0$, $B^0$, $D_s$, and $B_s$, and provided predicted
results for many unobserved energy levels in Table~\ref{d0}.
Although the use of a semirelativistic model may seem questionable for a system
containing a light quark, ultimately such an approach should be judged on
the basis of its predictions.
Additional experimental data on the light-heavy quarkonia should
be available in the near future.

	We have also used our results to test the accuracy of the heavy quark
effective theory.  According to Table~\ref{d0}, the heavy quark
expansion with the inclusion of the $M^{-1}$ and $M^{-1}\ln M$ corrections
yields generally good results for the $B^0$ and $B_s$ energy levels.
It does not, however, provide equally good results for the
energy levels of $D^0$ and $D_s$, which
indicates that the effective theory can be applied more
accurately to the $b$ quark than the $c$ quark.
We further find that the results for the energy levels in the
limit $M\to\infty$ are unacceptable.

        This work was supported in part by the U.S. Department of Energy
under Grant No.~DE-FG02-85ER40209.

\bibliographystyle{unsrt}

\begin{table}[b]
\caption{\label{d0} $D^0$, $D_s$, $B^0$ and $B_s$ energy levels
in MeV.  Effective theory results are given with the $M^{-1}$ and $M^{-1}\ln M$
corrections as well as in the limit of $M\to\infty$.  }
\bigskip
\centerline{ \begin{tabular}{lr@{$\pm$}lccc}
\hline\hline
\hspace*{1pt}&\multicolumn{2}{c}{Expt.}&Theory&Effective theory&$M\to\infty$\\
\hline
$1\;{}^1S_0\,\ (D^0)$&	1864.5&0.5	&	1864.5& 1864.5& 1864.5\\
$1\;{}^3S_1\,\ (D^{\star 0})$&2007&1.4	&	2007.0& 2010.9& 1864.5\\
$2\;{}^1S_0\, $&	\multicolumn{2}{c}{}	&	2547.7& 2566.5& 2431.9\\
$2\;{}^3S_1\, $&	\multicolumn{2}{c}{}	&	2647.0& 2662.1& 2431.9\\
$1\;{}^3P_0\, $&  \multicolumn{2}{c}{}	&	2278.6& 2310.2& 2244.8\\
$1\;{}^3P^\prime_1\, $& \multicolumn{2}{c}{} &	2407.3& 2414.6& 2244.8\\
$1\;{}^3P_2\,\ (D^0_2)$&	2465&4.2	&	2465.0& 2474.0& 2287.2\\
$1\;{}^1P^\prime_1\,\ (D^0_1)$&2421&2.8	&	2421.0& 2438.2& 2287.2\\
$\theta$&	\multicolumn{2}{c}{}	&	29.0$^\circ$&	30.9$^\circ$&	35.6$^\circ$\\
\hline
$1\;{}^1S_0\,\ (D_s)$&	1968.8&0.7	&	1968.8& 1968.8& 1968.8\\
$1\;{}^3S_1\,\ (D^\star_s)$&2110.3&2.0	&	2110.5& 2113.1& 1968.8\\
$2\;{}^1S_0\, $&	\multicolumn{2}{c}{}	&	2656.5& 2678.8& 2536.5\\
$2\;{}^3S_1\, $&	\multicolumn{2}{c}{}	&	2757.8& 2774.3& 2536.5\\
$1\;{}^3P_0\, $&  \multicolumn{2}{c}{}	&	2387.8& 2422.2& 2382.2\\
$1\;{}^3P^\prime_1\, $& \multicolumn{2}{c}{} &	2521.2& 2528.8& 2382.2\\
$1\;{}^3P_2\,\ (D_{s2})$&	2573.2&1.9	&	2573.1& 2582.8& 2402.8\\
$1\;{}^1P^\prime_1\,\ (D_{s1})$&2536.5&0.8	&	2536.5& 2552.1& 2402.8\\
$\theta$&	\multicolumn{2}{c}{}	&	26.0$^\circ$&	31.8$^\circ$&	35.6$^\circ$\\
\hline
$1\;{}^1S_0\,\ (B^0)$&	5278.7&2.1	&	5278.7& 5278.7& 5278.7\\
$1\;{}^3S_1\,\ (B^{\star 0})$&5324.6&2.1	&	5324.0& 5325.8& 5278.7\\
$2\;{}^1S_0\, $&	\multicolumn{2}{c}{}	&	5892.1& 5893.9& 5846.3\\
$2\;{}^3S_1\, $&	\multicolumn{2}{c}{}	&	5924.3& 5927.1& 5846.3\\
$1\;{}^3P_0\, $& \multicolumn{2}{c}{}	&	5689.5& 5692.5& 5659.1\\
$1\;{}^3P^\prime_1\, $& \multicolumn{2}{c}{} &	5730.8& 5734.1& 5659.1\\
$1\;{}^3P_2\, $&	\multicolumn{2}{c}{}	&	5759.1& 5761.4& 5701.5\\
$1\;{}^1P^\prime_1\, $& \multicolumn{2}{c}{}&	5743.6& 5745.4& 5701.5\\
$\theta$&	\multicolumn{2}{c}{}	&	31.7$^\circ$&	31.3$^\circ$&	35.6$^\circ$\\
\hline
$1\;{}^1S_0\,\ (B_s)$&	5383.3&6.7	&	5383.3& 5383.3& 5383.3\\
$1\;{}^3S_1\,\ (B_s^\star)$&5430.5&2.6	&	5431.9& 5434.1& 5383.3\\
$2\;{}^1S_0\, $&	\multicolumn{2}{c}{}	&	6000.9& 6003.1& 5950.9\\
$2\;{}^3S_1\, $&	\multicolumn{2}{c}{}	&	6035.8& 6039.1& 5950.9\\
$1\;{}^3P_0\, $& \multicolumn{2}{c}{}	&	5810.1& 5814.2& 5796.7\\
$1\;{}^3P^\prime_1\, $& \multicolumn{2}{c}{} &	5855.0& 5857.9& 5796.7\\
$1\;{}^3P_2\, $&	\multicolumn{2}{c}{}	&	5875.2& 5878.1& 5817.1\\
$1\;{}^1P^\prime_1\, $& \multicolumn{2}{c}{}&	5860.2& 5863.2& 5817.1\\
$\theta$&	\multicolumn{2}{c}{}	&	27.3$^\circ$&	27.1$^\circ$&	35.6$^\circ$\\
\hline\hline\end{tabular} }
\end{table}
\end{document}